\documentclass[english,11pt]{article}
\usepackage{geometry}
\usepackage{float}
\usepackage{pgf,tikz}
\usepackage{mathrsfs}
\usetikzlibrary{arrows}
\usepackage{mathrsfs,bm}
\usepackage{amsmath}
\usepackage{amssymb}
\usepackage{fancyhdr}
\usepackage{epsfig}
\usepackage{amsthm}
\usepackage{mathrsfs}
\usepackage{amsfonts}
\usepackage{mathrsfs}
\usepackage{enumerate}
\usepackage{subfigure}
\usepackage{setspace}

\usepackage{showlabels}
\usepackage{mathtools}
\usepackage{tikz}
\usepackage{tikz-3dplot} 
\usepackage{pgfplots}

\usepackage{amsmath,amssymb}
\usepackage{caption}
\usepackage{color}% Include colors for document elements
\usepackage{dcolumn}% Align table columns on decimal point
\usepackage{bm}% bold math
\usepackage{tikz}
\usepackage{tikz-3dplot} 
\usepackage{pgfplots}
\pgfplotsset{compat=1.11}
\usepackage{algorithm}
\usepackage[noend]{algpseudocode}
\usepackage{tikz}
\usetikzlibrary{bayesnet}
\usepackage{amssymb}
\usepackage{caption}
\usepackage{color}% Include colors for document elements
\usepackage{dcolumn}% Align table columns on decimal point
\usepackage{etoolbox}

\usepackage{amsmath}
\usepackage[noend]{algpseudocode}

\usepackage[utf8]{inputenc}

\usepackage[final]{hyperref} % adds hyper links inside the generated pdf file
\usepackage{setspace}

\usepackage[]{biblatex}
 
\hypersetup{
	colorlinks=true,       % false: boxed links; true: colored 
	linkcolor=blue,        % color of internal links
	citecolor=blue,        % color of links to bibliography
	filecolor=magenta,     % color of file links
	urlcolor=blue         
}
\usepackage{breqn}

\title{Robustness on Exchangeable networks}
\author{Marios Papamichalis \thanks{Purdue University, Postdoctoral Researcher at the Statistics Department, mpapamic@purdue.edu.}, Sim\'on Lunag\'omez \thanks{Lancaster University, Department of Mathematics and Statistics at Lancaster University, s.lunagomez@lancaster.ac.uk}, Patrick J. Wolfe \thanks{Purdue University, Frederick L. Hovde Dean of the College of Science and Miller Family Professor of Statistics, patrick@purdue.edu.} }

\begin{document}

\maketitle

\begin{abstract}
We adopt the statistical framework on robustness proposed by Watson and Holmes in 2016 and then tackle the practical challenges that hinder its applicability to network models. The goal is to evaluate how the quality of an inference for a  network feature degrades when the assumed model is misspecified. Decision theory methods aimed to identify model missespecification are applied in the context of network data with the goal of investigating the stability of optimal actions to perturbations to the assumed model. Here the modified versions of the model are contained within a well defined neighborhood of model space. Our main challenge is to combine stochastic optimization and graph limits tools to explore the model space. As a result, a method for robustness on exchangeable random networks is developed. Our approach is inspired by recent developments in the context of robustness and recent works in the robust control, macroeconomics and financial mathematics literature and more specifically and is based on the concept of graphon approximation through its empirical graphon.   \\


\end{abstract}

\newpage

\section{Introduction}

In this paper, we propose an approach for assessing the robustness of a Bayesian inference on a network model. We do this in the context of node-exchangeable network models. We adopt the decision theory perspective on Bayesian inference, where the summary to be reported from the posterior is obtained by optimizing the expected loss. The question that motivates our work can be phrase as follows: How does the quality of a Bayesian inference degrade when the assumed network model is misspecified? Here the quality of the inference refers to the value of the Bayes risk. We adopt the rationale that a small variation of the Bayes risk across a neighborhood of the assumed model indicates that the model in question is robust with respect to that inference.\\

Networks can be seen as an instance of high-dimensional data. They are particularly challenging to model because the structure of a single network can be described in terms of features that manifest at different scales (\emph{e.g.} motif counts, communities, degree distribution). Statistical models for networks often aim to describe all these features simultaneously, but in terms of inference, we are often interested in one feature in particular. In real world, random graphs are useful to understand stochastic processes that happen over a network, such as voter turnout (voter model), epidemics (SIS model) or how people's brains are wired differently, where most people are able to perform similar motor tasks, such as picking up an object. By generating random graphs we can control some conditions such as degree distribution, clustering coefficient, and degree sequence enables studying how real life network structures come to be. This often leads to the following situations: The feature we are interested in is not expressed explicitly as a function of the parameters of the model. e.g. inferring different types of centralities in a random network, which is described from degree distribution on epidemics, let's say cities or rural areas are the blocks of a Stochastic Block Model and the disease is prevalence of HIV. Moreover, one or more key assumptions of the model can be put into question. e.g. assumptions of conditional independence, assumptions regarding the degree distribution. \\

A principled approach for assessing the robustness of a Bayesian model under these circumstances is provided in \cite{Watson}. Their approach consists on evaluating the quality of a given inference when the assumed model is misspecified. To achieve this, a  set of possible generative models, that resemble the assumed model,  is defined. This set is called a \emph{neighbourhood} of the assumed model; its elements are not necessarily subject to exactly same set of assumptions as the assumed model. More precisely, their approach is based on the idea that the expected loss for the optimal action (\emph{i.e.}, the inference) under the assumed model should be roughly constant on a neighborhood of that model in order to call it robust with respect to that inference. However, the implementation of their framework to the context of network models entails the following challenges: How do we define a neighborhood in the space of models and how to we identify the worst case scenario in that neighbourhood? \\

The scientific theory behind robustness is well established and being used from the early 80s (e.g. \cite{Berger}) until the 1990s and 2000s when computational advances
and hierarchical models broadly outpaced the
complexity of data sets being considered by statisticians. In more recent times, very high-dimensional data are becoming common, the so called a big data era, whose size and complexities prohibit application of fully specified carefully crafted models. In \cite{Maccheroni,Berger,Hansen} review recent Bayesian decision theory research based on a local-minimax approach and in \cite{Wald,Vidakovic} Wald's minimax and
Savage's expected loss criterion are defined. Approaches based on tilted likelihoods, which we focus and use in this paper, can be
found in \cite{Miller,Grunwald} and mainly reviewed in \cite{Watson}. The theory of \cite{Watson} is recapped and extended here. Regarding networks, the literature of exchangeability of infinite binary arrays - which encompasses
networks - is very well understood: see for example \cite{Aldous1981,Hoover,Aldous1985},
\cite{Lauritzen1,Lauritzen2} and \cite{Kallenberg}. Of particular relevance here is the work
of \cite{Diaconis} (but see also \cite{Roy}), which details the
connections between exchangeability of random graphs and the notion of graph limits
developed in \cite{Lovasz}. In \cite{Patrick1}, \cite{Airoldi} and \cite{Latouche} the authors describe methods of how to approximate network limits. As an extend in \cite{Patrick2} the authors present the network histogram, a version of which we are will use in this paper.\\

%In this paper, we take advantage of a toolbox that includes the simulation from graphons to simulated annealing to put together a version of the framework proposed by \cite{Watson} so it can be implemented for node-exchangeable network models. \\

%The main contribution of our paper can be phrased as follows: First, we approach in a principled way the problem of assesing model misspecification of network data so that it overcomes the practical challenges of \cite{Watson} in terms of implemention; Second, we provide a Monte Carlo approach for modifying a given model based on graphons; this is one of the first applications of graphons to a practical statistical problem.\\

The core idea of our methodologies can be summarized as follows: We cast the ideas proposed by \cite{Watson} in the context of network data. The objective is not to provide additional theoretical results from those in \cite{Watson}, instead our focus is on developing tools that enable a statistician to implement that framework to exchangeable network models. Our contributions include the use of graphons for constructing a neighborhood around the assumed network model, the use of simulated annealing to obtain the a Monte Carlo approximation for the worst case scenario in that neighbourhood and finally the formulation of empirical experiments to explore the relationship between the size of the neighbourhood and the behaviour of diagnostics for Bayesian models.\\

%We consider the general question of how far a distribution, is from a specific centering distribution without adapting the \cite{Watson} decision making settings. Therefore, we provide
%some guidance on how to answer the above questions, resorting to bayesian diagnostic tools such as posterior predictive checks \cite{Gelman}. One could use the Bayesian $x^2$ test for goodness-of-fit of \cite{Valen}. Moreover, by using the most common
%measure of divergence between densities, the Kullback-Leibler (KL) divergence, we apply stochastic optimization algorithms in the space of non parametric models in order to find the model that maximizes the expected loss function in the model space. \\

The paper proceeds as follows:  In  Section 2,  we describe settings of the problem, we formulate them and give notation and definitions of exchangeable networks, exchangeable random networks, graphons and model space. Then, in section 3, we focus in our main purpose of this paper which is how to use and apply current tools on random networks to see how robust a random network is regarding the inference of a specific feature. Conceptually and computationally, our methodologies are presented. In section 4, for random network models data analysis that gives experimental results involving graphons, empirical graphons, simulated annealing,robustness and model misspecification is conducted showing the results of our approach. Finally, in section 5, we present with more details the future work involving overlapping research areas.

\section{Preliminaries}

Graphs in real life applications now days exist everywhere. Network data typically consist of a set of $n$ nodes and a relational tie random variable $G_{ij}$, measured on each possible ordered pair of actors, ($i,j$), $i$, $j$ = 1,$,\dots$, $n$, $G=(V, E)$ where $G$ is the network, $V$ the set of all vertices and $E$ the set of all edges that contain. In the simplest cases, $G_{ij}$ is indicating the presence or absence of some relation of interest, such as friendship, collaboration, transmission of information or disease, etc. The data are often represented by an $n \times n$ adjacency matrix $G$, with diagonal elements, representing self-ties, treated as structural zeros. In the case of binary relations, the data can also be thought of as a graph in which the nodes are actors and the edge set is {($i$,$j$): $G_{ij}$=1}. For many networks the
relations are undirected in the sense that $G_{ij}$=$G_{ji}$, $i$,$j=$1$,\dots$, $n$. In this project, we are dealing with undirected networks.\\

 Random graph is the general term to refer to probability distributions over graphs. Random graphs may be described simply by 
a probability distribution, or by a random process which generates them. The theory of random graphs lies at the intersection between graph theory and probability theory.\\

For this paper, we focus on node-exchangeable models for networks. Formally, an exchangeable sequence of random variables is a finite or infinite sequence $X_1, X_2, X_3, \dots$ of random variables such that for any finite permutation $\sigma$ of the indices 1, 2, 3, $\dots$, (the permutation acts on only finitely many indices, with the rest fixed), the joint probability distribution of the permuted sequence
$X_{\sigma (1)},X_{\sigma (2)},X_{\sigma (3)},\dots$ is the same as the joint probability distribution of the original sequence. In networks, this means that the distribution on $G= (V,E)$ is invariant on the indexing of $V$. A common feature shared by many network models is that of invariance to the relabeling of the network units, or (finite) exchangeability, whereby isomorphic graphs have
the same probabilities, and are therefore regarded as statistically equivalent. Exchangeability is a basic form of probabilistic invariance, but also a natural and convenient simplifying assumption to impose when formalizing statistical models for random graphs.\\

In the probability literature, node exchangeability of infinite binary arrays is very well understood: see for example \cite{Aldous1981, Aldous1985}. Of particular relevance here is the work
of \cite{Diaconis}, which details the
connections between exchangeability of random graphs and the notion of graph limits
developed in \cite{Lovasz}.\\

Examples of popular network models which rely on exchangeability include many exponential random graph models, the stochastic block model, graphon-based models, latent space models, to name a few.
\subsection{Graphons}

A graphon is a symmetric measurable function ${\displaystyle W:[0,1]^{2}\to [0,1]}$. Usually it is  understood as defining an exchangeable random graph model according to the following scheme: Each vertex $j$ of the graph is assigned an independent random value ${\displaystyle u_{j}\sim U[0,1]}$. Edge $(i,j)$ is independently included in the graph with probability ${\displaystyle W(u_{i},u_{j})}$.\\

A random graph model is an exchangeable random graph model if and only if it can be defined in terms of a (possibly random) graphon in this way. The model based off a fixed graphon $W$ is sometimes denoted ${\displaystyle \mathbb {G} (n,W)}$, by analogy with the Erd{\"o}s-R{\'e}nyi model of random graphs. A graph generated from a graphon $W$ in this way is called a $W$-random graph. It follows from this definition and the law of large numbers that, if ${\displaystyle W\neq 0}$, exchangeable random graph models are dense almost surely. The simplest example of a graphon is ${\displaystyle W(x,y)\equiv p}$ for some constant $p \in [0,1]$. In this case the associated exchangeable random graph model is the Erd{\"o}s-R{\'e}nyi model $G(n,p)$ that includes each edge independently with probability $p$.\\

If we instead start with a graphon that is piecewise constant by: dividing the unit square into $k\times k$ blocks, and
setting $W$ equal ${\displaystyle p_{lm}}$ on the ${\displaystyle (l,m)^{\text{th}}}$ block,
the resulting exchangeable random graph model is the $k$ community stochastic block model, a generalization of the Erd{\"o}s-R{\'e}nyi model. We can interpret this as a random graph model consisting of $k$ distinct Erd{\"o}s-R{\'e}nyi graphs with parameters ${\displaystyle p_{ll}}$ respectively, with bigraphs between them where each possible edge between blocks ${\displaystyle (l,l)}$ and ${\displaystyle (m,m)}$ is included independently with probability ${\displaystyle p_{lm}}$. When $k\times k$ blocks are equal in size then we have an empirical graphon. As mentioned implicitly before, as $k$ grows empirical graphon converges to the graphon. Specifically:\\

{\bf Definition:} A graph on $n$ vertices can be represented as a piecewise constant graphon by dividing
the unit square $[0, 1]^2$
into $n^2$
small squares of side length $1/n$, and representing the presence of
an edge between vertices $i$ and $j$ by a 1 on the squares $ij$ and $ji$, and its absence by a 0. This
representation is called the empirical graphon of the graph.

\section{Methodology}

Graphs in real life applications now days exist everywhere. Network data typically consist of a set of $n$ nodes and a relational tie random variable $G_{ij}$, measured on each possible ordered pair of actors, ($i,j$), $i$, $j$ = 1,$,\dots$, $n$, $G=(V, E)$ where $G$ is the network, $V$ the set of all vertices and $E$ the set of all edges that contain. In the simplest cases, $G_{ij}$ is indicating the presence or absence of some relation of interest, such as friendship, collaboration, transmission of information or disease, etc. The data are often represented by an $n \times n$ adjacency matrix $G$, with diagonal elements, representing self-ties, treated as structural zeros. In the case of binary relations, the data can also be thought of as a graph in which the nodes are actors and the edge set is {($i$,$j$): $G_{ij}$=1}. For many networks the
relations are undirected in the sense that $G_{ij}$=$G_{ji}$, $i$,$j=$1$,\dots$, $n$. In this project, we are dealing with undirected networks.\\

Algorithm 1, below, describe explicitly all the steps we follow:

\begin{algorithm}[H] 
 \caption{Algorithm for Robustness}
\begin{enumerate}
\item Obtain the graphons of the approximated model. 
\item Discretize the graphon, by using an $n\times n$ grid, get a SBMs and compute the KL in order to find the radius $C$ of the sphere which includes all the exchangeable random network models (fulfilling the assumption made). If the generating model is not known then we try to guess the value of $C$ by starting from a low value and consistently increasing it for each experiment.
\item Calibrate the parameter $T$ of the simulated annealing.
\item The function $h$ of the simulated annealing is the expected loss for the inference a specific graph feature .
\item Explore the models inside the ball using the perturbing and rescaling moves and find the maximum expected loss of that specific feature using simulated annealing.
\end{enumerate}
\end{algorithm}

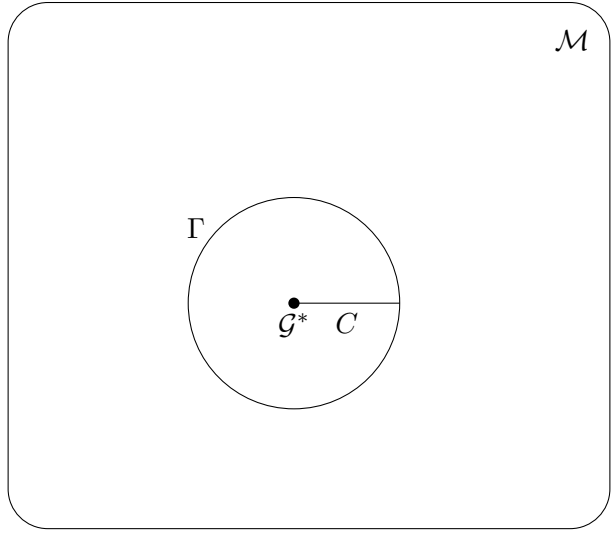
\begin{figure}
\begin{center}
\begin{tikzpicture}
  \draw [rounded corners=15pt] (3,2) rectangle ++(8,7);
\draw[] (10.5,8.5) node {$\mathcal{M}$};
\draw[] (5.5,6) node {$\Gamma$};
  \node at (6.80,5) {
\begin{tikzpicture}
% the origin
\coordinate (O) at (0,0);
\draw (O) node[circle,inner sep=1.5pt,fill] {} circle [radius=40pt];
\draw 
  (40pt,0) coordinate (xcoord) -- 
  node[midway,below] {$C$} (O) ;
\draw[] node[below] {$\mathcal{G}^{*}$};
\end{tikzpicture}
  };
\end{tikzpicture}
\end{center}
\caption{Space of Models $\mathcal{M}$. $\Gamma$ is the subspace ball in which we examine if the robustness of the model in the center. $C$ is the radius of the ball and $G^{*}$ the center model.}
\end{figure}

To follow the reasoning used in \cite{Watson}, we first define the space of modified versions of the approximating model. In the context of this paper, this is given by node-exchangeable graph models that are close to the approximating model in terms of Kullback-Liebler divergence. We take advantage of the fact that node exchangeable random network models can be represented non-parametrically, as graphons. We follow the rationale that similar graphons, based on a distance in function space, lead to similar models. Similarity is encoded by the Kullback-Liebler divergence. We introduce some additional notation: the model which generates the data is denoted by $\mathcal{G}$ and the approximating model $\mathcal{G}^{*}$. An assumption of this approach is that, the ball that contains all the modified versions of the approximated model we want to consider, also contains the model that generated the data. We denote the radius of such ball by $C$. The Kullback-Leibler is a pre-metric and it generates a topology on the space of probability distributions. We define the closed ball with center $\mathcal{G}^{*}$ and radius $C$ such that:
\begin{equation}
\Gamma(\mathcal{G}^{*}) = \{\mathcal{G} : {\text KL}(\mathcal{G}^{*},\mathcal{G})\leq C\}
\end{equation}
One example of a neighborhood is given by $C$ is the $\epsilon$ - contamination neighborhood from \cite{Berger2} formed by the mixture model 
\begin{equation}
\Gamma = {\mathcal{G}_{current} = (1 - \epsilon) \mathcal{G}^{*} + \epsilon q, q \in Q},
\end{equation}
where $\epsilon$ is the perceived contamination error, which in our case is provided by two moves, in $\mathcal{G}^{*}$ and $\mathcal{Q}$ is a class of contaminant distributions. \\

The next step is to perform a exploration-exploitation procedure to the ball with radius $C$ centered at the approximating model. In the context where the approximating model is given by a node-exchangeable network model, we need to first compute its graphon. In the case of the Erd{\"o}s-R{\'e}nyi model, the graphon is a constant function; for the SBM, the graphon is a piece-wise constant function. In general, the graphon can be discretized as in \cite{Patrick1, Airoldi, Latouche} by splitting the unit square in a large number ($n^2$ of equal squared cells). In order to move the model space defined by the ball with radius $C$, we need to shuffle and change the heights of the grid. To ease the exposition, we regard the grid as fixed, with as we mentioned $n\times n$ equal cells, which give the approximated SBM of a graphon. We provide two ways to explore the robustness of the models. Both of them rely on Simulated annealing. The first one, uses two kernels in order to choose the next model inside the ball the the second approach relies on \cite{watson2017characterizing} where the authors characterise variation of expected loss within the neighbourhood $C$. 

\subsection{First approach}

We follow two moves to explore the ball, are the perturbing move and the rescaling move; these are described as follows:

\begin{itemize}
\item In order to perturb the discretized graphon and move in models inside the ball that are defined by KL between the approximate model and the perturbed models like in \cite{Watson}, we use a simplex like below:
\begin{equation}
\{K\in \mathbb{R} :K_{1,1}+\dots +K_{n,n}=\alpha,K_{i,j}\geq 0,i,j=0,\dots ,n\}
\end{equation}
Initially, parameter $\alpha$ above is 1 but changes its value less than 1 due to the second move. By changing the probabilities of $K_{i}$, we perturb each current model every time. (assumptions: same sized squares, graphon is symmetric and simplex). For this:

Draw $n^2$ independent random samples $y_{1,1},\dots,y_{n,n}$ from Gamma distributions each with density:\\
\begin{equation}
\textrm{Gamma}(\alpha_{i,j}, 1) = \frac{y_{i,j}^{\alpha_{i,j}-1} \; e^{-y_{i,j}}}{\Gamma (\alpha_{i,j})}
\end{equation}
where , $\alpha_{i,j}$ which denotes the counts in each cell, and then set
\begin{equation}
K_{i,j} = \frac{y_{i,j}}{\sum_{i,j=1}^{n} y_{i,j}}
\end{equation}
If $y_{i,j}$ are independent $\mathrm{Gamma}(\alpha_{i,j},1)$, for $i,j=1,\dots,n$ then:
\begin{equation}
(K_{1,1},\dots,K_{n,n}) = \left
(\frac{y_{1,1}}{\sum_{i,j=1}^{n} y_{i,j}}, \dots, \frac{y_{n,n}}{\sum_{i,j=1}^{n} y_{i,j}} \right) \sim \mathrm{Dirichlet}(\alpha_{1,1},\dots,\alpha_{n,n}) \, .
\end{equation}

\item Scaling the graphon: E.g. for the Erd{\"o}s-R{\'e}nyi model the initial graphon, which has equal degree density and is flat, will be equally to one. Then, to get the desired surface we multiply the empirical graphon heights, that approximate the graphon, with $\rho$. When we scale it, it is no longer a graphon. It is a scaled graphon. In our case $\rho$ is randomized with $1\pm
\delta$, where $\delta$ has a small value that varies.
\end{itemize}

The modified model that results from applying either of those moves at any given iteration has to be contained inside the ball of radius $C$.\\

\begin{figure}
\begin{center}
\begin{tikzpicture}
  \draw [rounded corners=15pt] (3,2) rectangle ++(8,7);
\draw[] (10.5,8.5) node {$\mathcal{M}$};
  \node at (7,5.5) {
\begin{tikzpicture}
% the origin

\coordinate (O) at (0,0);
\draw (O) node[circle,inner sep=1.5pt,fill] {} circle [radius=90pt];
\draw[] node[below] {$\mathcal{G}^{*}$};
\draw [line width=0.5mm, red ] (2,1.3) -- (2,1.3) node[right] {$x$};
\draw [line width=0.5mm, red ] (2,1.3) -- (1.7,0.9) ;
\draw [line width=0.5mm, red ] (1.7,0.9) -- (1.6,1.5);
\draw [line width=0.5mm, red ] (1.6,1.5) -- (1,1.2) node[above] {Perturbing};
\draw [line width=0.5mm, red ] (1,1.2) -- (1.4,0.5);
\draw [line width=0.5mm, red ] (1.4,0.5) -- (1.9,0.8);
\draw [line width=0.5mm, red ] (1.9,0.8) -- (1.9,0.8) node[right] {$y$};
\draw [line width=0.5mm, blue ] (1.9,0.8) -- (1.4,-0.9) node[midway,right] {Rescaling};
\draw [line width=0.5mm, blue ] (1.4,-0.9) -- (1.4,-0.9) node[above] {$z$};
\draw[] node[below] {$\mathcal{G}^{*}$};
\draw [] (1.5,1) ellipse [x radius =1, y radius = 0.5];
\node [at = {(1.5,1)}] {$A$};
\draw [] (1.5,-1) ellipse [x radius =1, y radius = 0.2];
\node [at = {(1.5,-1)}] {$B$};
\end{tikzpicture}
  };
\end{tikzpicture}
\end{center}
\caption{Exploring and exploiting the Models in the ball. The red trajectory inside a subspace $A$ illustrates the perturbing move which is created by changing the heights of two cells of the empirical graphon. With the blue line, the rescaling move is illustrated, jumping to another subspace $B$ of the ball.}
\end{figure}
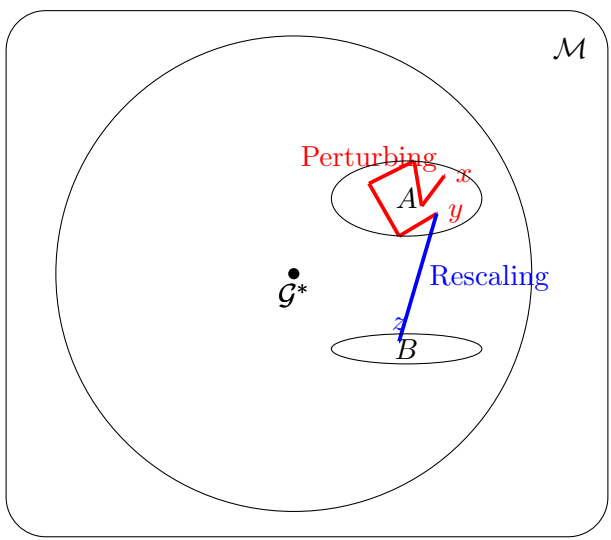
%So there are two graphons represented be two piecewise constant 3D SBMs.\\

The Kullback-Leibler between the graphon approximations of the two random graph models needs to be approximated: 
\begin{equation}\label{Eq:KullLieb}
\text{KL}(p \mid \mid q)=\int_{-\infty}^{\infty} \int_{-\infty}^{\infty} p(x, y) log\frac{p(x,y)}{q(x,y)} \text{d}x \text{d}y,
\end{equation}
without conditioning on the latent positions.\\
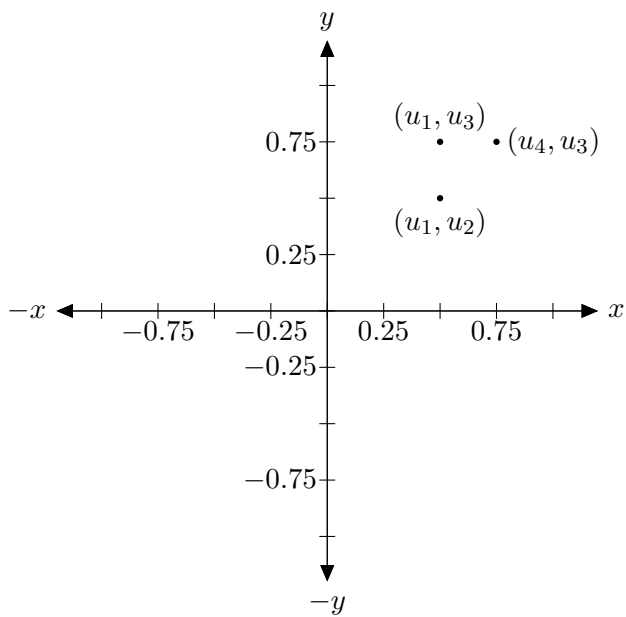
\begin{figure}
\begin{center}
\begin{tikzpicture}[scale=3]
\draw[thin,<->] (-1.2,0) -- (1.2,0) node[right] {$x$};
\draw[thin,<->] (0,-1.2) -- (0,1.2) node[above] {$y$};
\draw[thin,<->] (1.2,0) -- (-1.2,0) node[left] {$-x$};
\draw[thin,<->] (0,1.2) -- (0,-1.2)  node[below] {$-y$};

\foreach \x [count=\xi starting from 0] in {-1,-0.75,-0.5,-0.25,0,0.25,0.50,0.75,1}{% ticks
    \draw (\x,1pt) -- (\x,-1pt);
    \draw (1pt,\x) -- (-1pt,\x);
    \ifodd\xi
        \node[anchor=north] at (\x,0) {$\x$};
        \node[anchor=east] at (0,\x) {$\x$};
    \fi
}

\foreach \point in {(0.5,0.5),(0.75,0.75),(0.5,0.75)}{% points
        \fill \point circle (0.4pt);

}
	\node [below] at (0.5,0.5) {$(u_1,u_2)$};
	\node [right] at (0.75,0.75) {$(u_4,u_3)$};
	\node [above] at (0.5,0.75) {$(u_1,u_3)$};
\end{tikzpicture}

\end{center}
\caption{Latent positions projected in x,y-axis.}
\end{figure}

One computational challenge that arises due to the graphon representation is that some the latent positions, e.g $(u_1,u_2)$ and $(u_1, u_3)$, are dependent as they have the coordinate $u_1$ in common. This problem disappears once we condition on the values of $u_1,u_2$ and $u_3$. However, the computation in equation \ref{Eq:KullLieb} does not rely on conditioning on those variables. Observe that the entries of the adjacency matrix associated to $(u_1, u_2)$ and $(u_4,u_3)$ will be independent, without conditioning (Figure 3).\\ 

\begin{algorithm}[H]
 \caption{Monte Carlo algorithm for approximating Kullback-Leibler}
\begin{enumerate}
\item Draw a point pattern on $[0,1]$ by letting $u_i\sim \text{Unif}(0,1)$, then compute the pairs $(u_i,u_j)$ and then reflect with respect to the diagonal given by the identity.
\item Since that grid is fixed essentially we count, take those points that they are occupied by the point pattern get the $\text{KL}_i$ for those in the upper triangle and then add them. 
\item Repeat steps 1 and 2 $n$ times to obtain $\text{KL}_1, \dots, \text{KL}_n$.
\item Average all the empirical KL results to obtain an estimate of the Kullback-Liebler.
\end{enumerate}
\end{algorithm}

The Monte Carlo Algorithm 2 provides an approximated average of Kullback-Liebler divergences over their dependencies. It preserves the structure and averages out where the possible latent positions are located.\\ 

Another approach, which can be used as filtering model to decrease running time, is to calculate the KL between two models and check whether we are inside the ball is following the \cite{Degroot}:\\

{\bf Lemma}: We order the heights sequence of the cells in decreasing order ($x_1,x_2,\dots ,x_{n^2}$) where $x_1 \geq x_2 \geq \dots \geq x_{n^2}$ then if after the moves we have heights ($y_1,y_2,\dots,y_{n^2}$) again in decreasing order and ($x_1 \geq y_1,x_2 \geq y_2,\dots,x_{n^2} \geq y_{n^2}$) then due to exchangeability the sequence $x_i$ is more sufficient than the sequence $y_i$. \\

{\bf Corollary}: In rescaling move, when $\rho \leq 1$, the previous ordering was more sufficient than the new one.\\

By taking advantage of these results, we can construct a more efficient filter algorithm (Algorithm 3) which verifies whether the modified model implied by the move is inside the ball or not.

\begin{algorithm}[H]
 \caption{Analytical estimation of Kullback-Leibler}
\begin{enumerate}
\item Otherwise we keep as max the sequence where ($x_1 \geq y_1,x_2 \geq y_2,\dots y_i \geq x_i,\dots,x_{n^2} \geq y_{n^2}$) which is inside the ball. So merging $x_i$'s and $y_i$s in a way we are inside the ball. If we keep that max sequence the we do not have to calculate the KL every time but only when we find a sequence where $y_i \geq x_i$ than that. All $x_i$ and $y_i \leq$1. 
\item The first move can start with max graphon with equal heights that satisfy $KL<C$ which is the radius of the ball.
\end{enumerate}
\end{algorithm}

\subsection{Second approach}

Generalised Bayesian bootstrap offers an
efficient and computationally straightforward way to compute posterior distributions and estimates, enjoying some advantages over Markov chain techniques. We obtain the original Bayesian bootstrap proposed by \cite{rubin1981bayesian} sampling from a Dirichlet with parameter vector $\alpha_n p$ gives a generalised
version of this bootstrap procedure. Let us now consider our setting with the SBM. Here, as before $\pi_I$ is a discrete density with $n^n$ atoms in the center of the ball, where
$\{\xi_{1,1}, \dots , \xi_{n,n}\}$, i.e., $p_I(x) = \sum_{i=1}^n \sum_{j=1}^n p_{i,j} \delta_{\xi_{i,j}}(x)$, with $p_{i,j}$ > 0 for all $i,j = 1,\dots, n$ and $\sum_{i=1}^n \sum_{j=1}^n p_{i,j}=1$. Let $w = (w_{1,1},\dots, w_{n,n})$ be random weights such that $w_{i,i} \geq 0$
and $\sum_{i=1}^n w_i=1$ almost surely. Let $\pi$ be a random distribution defined as
a reweighing of the atoms of $\pi_I$ with the random weights $w$. In notation,
$f(x) = \sum_{i=1}^n \sum_{j=1} w_{i,j} \delta_{\xi_{i,j}}(x)$. For our case, $\alpha_n=n^2$. This procedure is the same as perturbing the weights of the cells of the SBM from $p_{i,j}$ to $w_{i,j}$ randomly. For simplicity we denote $p_i$ with $i \leq n^2$ and $w_i$ with $i \leq n^2$.\\ 

In order to perturb the discretized graphon and move in the models inside the ball, that are defined by KL between the approximate model and the perturbed models like in [2], we use a simplex like below:
\begin{equation}
\{w\in \mathbb{R} :w_1+\dots +w_n^2=1,w_i\geq 0,i,j=0,\dots ,n^2\}
\end{equation}
By changing the weights of $w_{i}$, we perturb each current model every time. (assumptions: same sized squares, graphon is symmetric and simplex). The modified model has to be contained inside the ball of radius $C$. The following proposition comes from using the theory created in \cite{watson2017characterizing} for SBMs:\\

{\bf Proposition 1:} Let $\pi$ be a “generalised Bayesian bootstrap” draw a SBM around $\pi_I$
with weights of cells of $\pi$, $w \sim Dir(\alpha_n p)$. Then the Kullback-Leibler divergence given in has mean:
\begin{equation}
    E\{KL(\pi \mid \mid \pi_I)\} =\sum_{i=1}^{n^2} w_i\{\psi_0(\alpha_n w_i+1)-\psi_0(\alpha_n+1)\}-H(p)
\end{equation}

where $H(p) = \sum_{i=1}^{n^2} w_i log w_i$ the entropy of the vector $p$, and the variance
given by
\begin{align*}
Var(KL(\pi \mid \mid \pi_I)) = \sum_{i=1}^{n^2}\{Var(w_i logw_i\}+ (log p_i)^2 Var(w_i)-2(log p_i) Cov(w_i log w_i, w_i)\\
+2\sum_{i<j}\{Cov(w_i log w_i, w_j log w_j) + (log p_i)(log p_j)Cov(w_i, w_j ) - 2(log p_j )Cov(w_i log w_i, w_j )\}
\end{align*}

This result explores properties of the Kullback-Leibler of draws from
some SBM models with respect to their centering distribution, something which, in closed form, was impossible in the first method. 

\subsection{Simulated annealing}

The objective function for the simulated annealing is the expected loss associated to the inference of interest (prediction, estimation of a graph feature) computed with respect to the different models inside the ball of radius $C$. For instance, for the quadratic loss, we have:
\begin{eqnarray}
E[(\hat{\tau}(\mathcal{G})-\tau(\mathcal{G}))^2] & = & E[(\hat{\tau}(\mathcal{G})-E[\hat{\tau}(\mathcal{G})])^2]+(E[\hat{\tau}(\mathcal{G})]-\tau(\mathcal{G}))^2 \nonumber\\
                                                 & = & V(\hat{\tau}(\mathcal{G}))+(E[\hat{\tau}(\mathcal{G})]-\tau(\mathcal{G}))^2.
\end{eqnarray}
Here,  $\tau(\mathcal{G})$ is the true value of the current model feature and $\hat{\tau}(\mathcal{G})$ is an estimator. For estimating features from exchangeable models which are not straightforward to relate to the parametrization at hand (\emph{e.g.} random networks with a given diameter), we extract the information about the feature included in the empirical graphon through the corresponding SBM of the model (each node-exchangeable model can be represented by a SBM). 

\begin{algorithm}[H]
 \caption{Selecting T parameter for Simulated Annealing}
\begin{enumerate}
\item Sample the ball regarding the expected losses of the feature, combined with values of the ball boundaries and approximated model $\mathcal{G}^{*}$.
\item Find the median of those samples.
\item Give $T_0$ this value.
\end{enumerate}
\end{algorithm}

Another challenge is to propose a method how to select the initial value of parameter $T$ from the simulated annealing, the stochastic optimization technique we use. This value has to be calibrated according to the values of the expected loss associated to the inference for that feature. We select $T_0$ by Algorithm 4. $T_0$ and the losses have to be in the same scale and in order to select $T_0$ we sample from the ball of radius $C$.\\

{\bf Theorem 1}: Given two arbitrary points, representing exchangeable models inside the ball, there is a positive probability to reach one from another, based both on the two moves above or generalized Bayesian bootstrap.\\

The theorem above, essentially, indicates that every point in the ball, which has an node exchangeable model, can be reached, a result that is not provided by \cite{Watson}.

\section{Simulation Studies}

In this section we discuss three simulation studies: the first one shows that the simulated annealing is a reliable tool for exploring the ball centered at the assumed model; the second one is designed to illustrate the behaviour of our method for two concrete random graph models, and the third is designed to find an indicative value for $C$. \\

For all the subsections we consider the following setup: given the generative random graph model and the assumed node-exchangeable model we represent the assumed model and its modifications via graphons; we take advantage of the results that state that graphons can be approximated by stochastic block models. So, Theorem 1 is practically evaluated. The assumed model and the set of modifications constitute a ball of radius $C$ defined in terms of Kullback-Liebler divergence.  \\

The simulation regimes are defined by the random graph model, the corresponding vector of parameters and the sample size. These regimes are displayed in Table 1. To infer the generating model we use the empirical graphon approach. The graph features we considered were: the density of the networks and number of communities for an Erd{\"o}s-R{\'e}nyi model and two empirical graphons. The first empirical graphon ($EG_1$) is selected a random empirical graphon after 100 iterations of moves of Erd{\"o}s-R{\'e}nyi simulations. To infer the second empirical graphon ($EG_2$) parameters we use \cite{Latouche1} approach, produce a SBM and use the empirical graphon as grid, by using the same notation for number of blocks ($K$), and inclusion probabilities ($\lambda, \epsilon$). When we implement our method we consider 1000 samples of the desired parameters $\theta_i$ from the models inside the sphere and 100 networks instances given the value of the parameters.\\

\begin{table}
\centering
\fbox{%
\begin{tabular}{| l  l  l |}
\hline
Approximating Exchangeable RGN & Parameter Specification & Features\\
\hline            
$ER$  &  $K$=1, $N=100$ (fig. 3) & Blocks and Density\\
\hline
$EG_1$    &   Point of ER after 100 moves (fig. 5)  & Blocks and Density \\
\hline
$EG_2$    &   $\lambda=0.5$, $\epsilon=0.5$, $N=100$  & Blocks and Density \\
\hline     
\end{tabular}}
\caption{Approximating Exchangeable Random graph models, parameter vectors and graph features considered for setting up simulation regimes.}
\end{table}

\subsection{Variability of Stochastic Optimization process and data sets}

The objective of this experiment is to verify whether the simulated annealing is able to reach the worst case scenario (\emph{i.e.}, to find the maximum value expected loss within the ball of radius $C$). To this end, we explore the space defined by the ball by exhausting as many as possible models inside that space. Theoretical results for the second method can be found in \cite{Watson}, though they are can not be used for practical implementation. For the regimes displayed in Table 1, we examine the variability of the simulated annealing by running it 100 times per regime and then we compute the mean and the variance compared with the ground truth. We regard as ground truth the output from a brute force algorithm that explores 1.000.000.000 models inside the ball.  \\

\begin{table}[H]
\centering
\fbox{%
\begin{tabular}{| l  l  l  l l l l|}
\hline
- & - & - & Method 1 & Method 1 & Method 2 & Method 2\\
\hline
Approx. Exch. RGN & Feature & Brute Force & Mean & Variance &Mean & Variance  \\
\hline            
$ER$  &Degree&  7.65817e-7 & 7.65828e-7  & 9.1742e-17 & 7.657237-7  & 9.27162e-17 \\
\hline
$EG_1$   &Degree &   7.65817e-7 &  7.65831e-7  & 9.1765e-17&  7.652716e-7  & 9.12736e-17 \\
\hline
$EG_2$  &Degree  &   0.000929188  &0.000929213 & 8.9826e-8 &0.00092982711 & 8.9726e-8\\
\hline               
$ER$ & Blocks & 50.2342& 50.2748 & 0.00281 & 50.26839 & 0.00281\\
\hline
$EG_1$  & Blocks  &   48.9342 & 48.8281& 0.00301&48.88271& 0.003018\\
\hline
$EG_2$ & Blocks   &   1826.2342  & 1827.1977& 1.28392& 1828.1723& 1.282887\\
\hline     
\end{tabular}}

\caption{Expected loss of worst case scenario with brute force (ground truth) compared with the mean of expected loss of the worst case scenario for the density and blocks of the three different models providing the variance of Expected loss of worst case scenario}
\end{table}

As expected, Table 2 confirms the reliability of our approaches since the expected losses of the exhausting method fall into the set of values where the results from simulated annealing concentrate. The estimated variances can be regarded as very small. By, comparing the two approaches we can see the following:

\begin{itemize}
    \item Both approaches reach the worst case scenario as we it is proven by theorem 1, as well. 
    \item Even if the Proposition 1 gives an analytical expression for the second approach, digamma and trigamma distributions are hard to be calculated and essentially their calculation in every step is an approximation. This constitutes the first method in some cases faster, even though it is entirely approximating procedure. 
\end{itemize}

\subsection{Empirical Results}

In this section, we present and discuss the results of the proposed method for each one of the three different settings presented. As discussed in Section 3, our approach associates a score to each regime, given the value of $C$. This score is given by the maximum expected loss associated to the inference to the feature. In Figure 4, we illustrate the three approximating models which are located in the center of the ball, for each of the regimes.

\begin{figure}[H]
\centering
\includegraphics[scale=1]{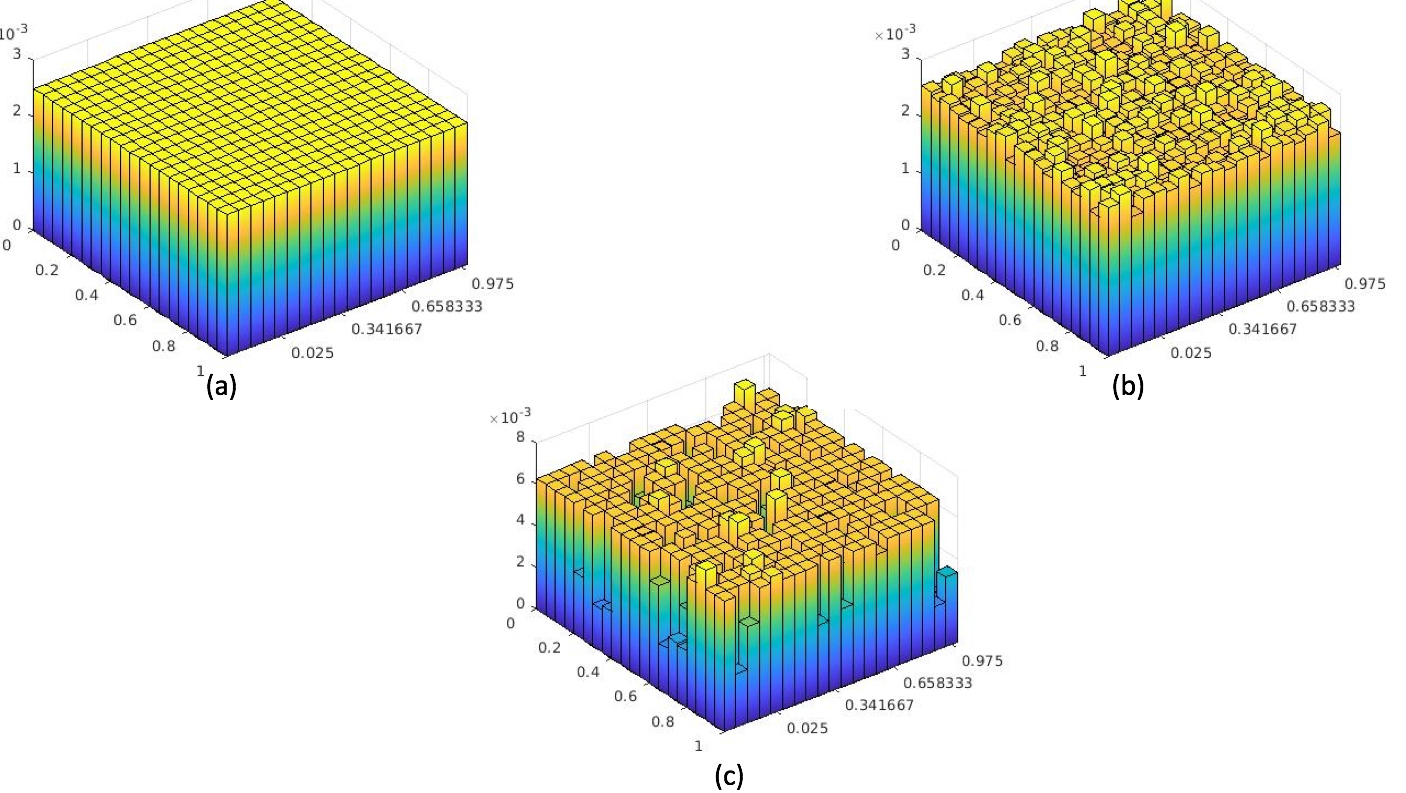}
\caption{(a) Erd{\"o}s-R{\'e}nyi model, 20 $\times$ 20 cells. (b) Empirical graphon, 20 $\times$ 20 cells, after applying the moves. This empirical graphon is used as an approximating model $EG_1$ for the second experimental design, as well. (c) Approximation of the graphon represented by an empirical graphon.}

\end{figure}

\begin{figure}[H]
\centering
\includegraphics[scale=0.4]{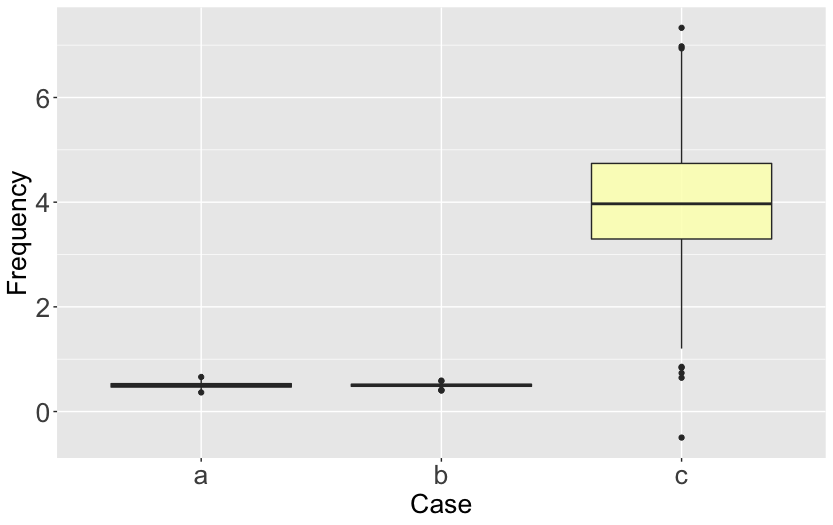}
\caption{Normalized values of frequencies (Expected Loss (model)-Expected Loss(center model))/Expected Loss(center model) for the density. Approximating model is 0 in all cases.}
\end{figure}

\begin{figure}[H]
\centering
\includegraphics[scale=0.4]{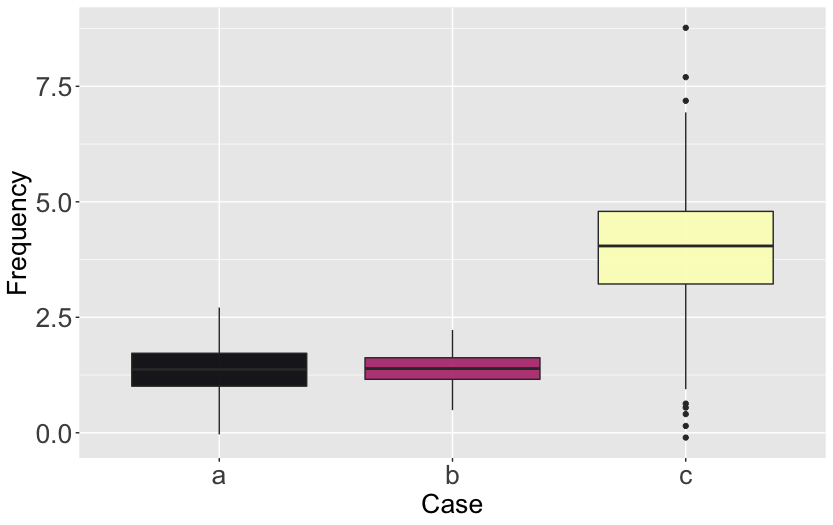}
\caption{Normalized values of frequencies (Expected Loss (model)-Expected Loss(center model))/Expected Loss(center model) for number of blocks in SBM. Approximating model is 0 in all cases.}
\end{figure}

\begin{table}[H]
\centering
\fbox{%
\begin{tabular}{| l  l  l l|}
\hline
Approximating RNM ($\mathcal{G}^{*})$ & Radius $C$ & A & B\\
\hline            
$ER$  & 0.921   & 2.03& 2.93\\
\hline
$EG_1$ &  0.921   & 1.94  &  2.01\\
\hline
$EG_2$    &  0.921    & 9.73 & 9.84\\
\hline       
\end{tabular}}
\caption{Results for three models. Radius $C$ given by Kullback-Leibler divergence is given and the maximum expected loss for one Erd{\"o}s-R{\'e}nyi model and two different empirical graphons are presented. Column A indicates the quantity $(Exp. Loss(model)-Exp. Loss(\mathcal{G}^{*}))/Exp. Loss(\mathcal{G}^{*}$) for the density and B indicates the quantity $(Exp. Loss(model)-Exp. Loss(\mathcal{G}^{*}))/Exp. Loss(\mathcal{G}^{*}$) for the community blocks. The fist two model are reasonable to be fitted but the last empirical graphon is not very robust in terms of inference for Density and Number of Communities.}
\end{table}
In Figures 5 and 6 we evaluate how robust the models in question are with respect to inferences on their density and community blocks. These results suggest that two of the models are robust in terms of the graph features used for the simulation. By this its is meant that the distribution of the scores concentrate around zero. On the other hand, the third model does not seem to be robust with respect to these features. The results show that a slight perturbation in the model might end up with a huge perturbation in terms of the values of the scores defined in Table 3. This criticism is complemented by the results presented in Table 3. 

\subsection{Relationship between lack of fit and radius}

The objective of this exercise is to provide intuition to the practitioner of how the size of the ball is related with a diagnostic, which indicates lack of fit. Until now, the radius of the ball was fixed and we were able to find the worst case scenario, which was compared with the assumed model at the center or of the ball. Intuitively, for different radii, the statistician should be able perform the same procedure and check whether the assumed model is robust or not. One way to explore the relationship between the radius and a diagnostic is to construct a sequence of models that are further away will have not a problematic value of the diagnostic compared with the worst model of larger balls. \\

For example, considering an Erd{\"o}s-R{\'e}nyi model as an approximating model and perform inference about average degree distribution, a posterior predictive check about this feature will tell us how the robustness of the model with respect to that feature is going to be affected. To do so, we consider an Erd{\"o}s-R{\'e}nyi model or a empirical graphon and construct the sequence of models by either increasing the number of cells or making the probability of inclusion more assortative (in line with SBM). \\ 

A different way to proceed would be to perform a binary search starting from a fixed radius and approximating the minimum robust ball. A first attempt on this direction would be to use a bisection-type search.\\%By this, it is meant that we would set $C$ to an arbitrary number $c$ and in case approximating model in the ball with radius ${\displaystyle c}$ is not robust due to the diagnostic then we devide $c$ by 2, else ${\displaystyle c}$ we multiply $c$ by 2 We repeat the procedure for $k$ times. \\

As mentioned before, the frequency of lack of fit should be happening more and more often as the radius increases. We use the same example, with the Erd{\"o}s-R{\'e}nyi model in the center of the ball, as in Subsection 4.2 using posterior predictive checks as a diagnostic and altering the radius $k$=100 times (Table 4).

\begin{table}[H]
\centering
\fbox{%
\begin{tabular}{| l  l  l |}
\hline
Initial Model &  Feature  & Approximated Radius $C$ \\
\hline            
Erd{\"o}s-R{\'e}nyi & Average degree distribution  & 1.592\\
\hline     
Erd{\"o}s-R{\'e}nyi & Blocks  & 10.743\\
\hline  
$EG_1$& Average degree distribution  & 1.404\\
\hline  
$EG_1$& Blocks  & 8.968\\
\hline  
$EG_2$& Average degree distribution  & 0.2817\\
\hline  
$EG_2$ & Blocks   & 1.873\\
\hline    
\end{tabular}}
\caption{Approximated minimum radius for Erd{\"o}s-R{\'e}nyi, $EG_1$ and $EG_2$ models for Average degree distribution and Blocks. }
\end{table}
\section{Discussion}

In this paper, we present two principled approaches for robustness on exchangeable random network models. To the best of our knowledge, this is one of the first efforts in the literature, where graphons are used to deal with practical problems. The main advantage of our method is that it provides the statistician with a conceptual frameworks that enables him\slash her to check whether an specific inference of an approximating model is robust with respect to model misspecification. Moreover, the two moves we presented allow us to visit every model in the model using simulated annealing together with the methodology of characterising variation of expected loss within the neighbourhood provided in \cite{Watson}.\\
 
One of the main challenges to develop these ideas is to provide a method to find the size of the neighborhood in which the model we try to fit is centered. In the literature, we are provided by the size of the neighborhood and is assumed a priori that the worst case scenario exists inside of it. This invokes the question of what should an appropriate size of this ball be? While the simulation studies provide some insight about this problem, a more principled approach is needed. In \cite{Watson}, the authors do not provide any answer for finding a size of the ball, which includes only reasonable models in case we have a goodness of fit
of Bayesian Models. Combining the logic of \cite{Murray} can provide us with a sense of how far is the data generating mechanism from the model we want to use in the first place and interpret.
\\

From our perspective, the main limitations of our work are: First, we do not provide a general way to represent a flexible way to perturb non-exchangeable models. Here, we used graphons as a tool for representing perturbed versions of the approximating model. In \cite{Caron, veitch1, veitch2, Borgs} we see that a graphon arise from a more general nonparametric object called graphex, a limit for measure exchangeable random networks. By substituting empirical graphon with empirical graphex, which comes from graphon processes, we can generalize our results. Second, we are dealing with the bias of estimators combining network models and network realizations for specific features e.g. the average distance. We need reliable estimators to calculate the generating model specific feature of the random network vs the properties of the network in order to estimate the expected loss. Otherwise, there might be a misspecification on the model. Third, we do provide methodologies to express and perturb the graphon and the model we are interested via computational and approximation approaches which come with a (small) error. \\

For future work we propose the following four directions: The first two are aimed to address the limitations mentioned above. First, we will look for estimators that enable us to connect random networks with networks in terms of features and properties, respectively. The second direction is extending our method to non-exchangeable models  \cite{Caron, veitch1, veitch2, Borgs}. Another way is to define and use flexible family of models and compute the distance with the approximating model \cite{Murray}. To induce perturbations we have to move from a parametric space to a non-parametric space, maybe observable space and then apply \cite{Murray} techniques. The third direction is to develop a method that can be applied to networks and (non) exchangeable random network models adopting the robustness setting in \cite{Patrick3}. The last direction is to apply our framework on coarsen network data as in \cite{heitjan1991ignorability}. This is of a specific scientific interest, since will provide the practitioner with a tool to investigate the behavior of coarse and missing data network problems like in \cite{heckathorn1997respondent} and \cite{lunagomez2018evaluating}.\\

\section*{Appendix}

{\bf Simulated Annealing}\\

Simulated annealing (SA) is a probabilistic optimization technique for approximating the global optimum of a given function. Therefore, rescaling
partially avoids the trapping attraction of local maximum. Given a temperature
parameter $T > 0$, a sample $\theta_1^T, \theta_2^T,\dots$ is generated from the distribution:
\begin{equation}
\pi(\theta) \propto \exp(h(\theta)/T)
\end{equation}
and can be used to come up with an approximate maximum
of $h$. As $T$ decreases toward 0, the values simulated from this distribution
become concentrated in a narrower and narrower neighborhood of the local maxima of $h$.
\begin{algorithm}[H]
 \caption{Simulated Annealing}
\begin{enumerate}
\item Simulate $\zeta$ from the distribution of $\pi(\theta)$.
\item Accept $\theta_{i+1}=\zeta$ with probability $\rho_i=\exp(\Delta h_i/T_i) \wedge 1$ where $\Delta h=h(\zeta)-h(\theta_i)$ for $i=0 \dots I$, $I$ the number of iterations and 
\[
 \theta_i=\begin{cases}
               \zeta\text{ with probability }\rho= \exp(\Delta h/T) \wedge 1\\
               \text{0 with probability }1-\rho
\end{cases}
\]
\item Update $T_i$ to $T_{i+1}$.
\end{enumerate}
\end{algorithm}
Therefore, if $h(\zeta) \geq h(\theta_0)$, $\zeta$ is accepted with
probability 1; that is, $\theta_0$ is always changed into $\zeta$. On the other hand, if
$h(\zeta) \leq h(\theta_0)$, $\zeta$ is still be accepted with probability $p \neq 0$ and $\theta_0$ is then
changed into $\zeta$. This property allows the algorithm to escape the attraction
of $\theta_0$ if $\theta_0$ is a local maximum of $h$, with a probability which depends on the
choice of the scale $T$, compared with the range of the distribution of $\pi$.\\

\subsection*{Scalability}

This task is high dimensional (NP-complete) and in order to make it scalable we divide the SBM into $n \times n$ cells of 5-6 nodes each. For each cell, we can sample all network realisations (observables) using sampling for constant degree, in the same spirit with the concordance function described below, as in \cite{Speed}:

\begin{equation}
P(G) \propto \exp\{\lambda \sum_{i=1}^{n^2} w_i f_i(G)\}=\prod_{i=1}^{n^2} \exp \{ \lambda w_i f_i(G) \}
\end{equation}

, where the $\exp\{\lambda w_i f_i(G)\}$ is one cell of one Erd{\"o}s-R{\'e}nyi model part of the whole SBM.\\ 

Then, we merge all the networks into one each time, extracting the feature (parameter) (\cite{Ying}). For this, we use concensus Monte Carlo (\cite{Scott}), in order to merge the observables for each feature. Obviously, this approach is computationally expensive, though it is scalable, due to SBM division in $n \times n$ squared cells (number of cores needed in Map-Reduce framework). \\

{\bf Lemma Proof:}\\

For every two cells accordingly in the two sequences, project to edges (range) $\binom{n}{2} \times p$ the following edges (domain) $\binom{n}{2} \times (p+p_1)$, where $p \leq 1$ and $p+p_1 \leq 1$. From domain, edges $\binom{n}{2} \times (p_1)$ are not mapped.\\

{\bf Proof of Proposition 1:}\\

For $i,j \in \{1,\dots,n^2\}$, note that each $w_i \sim Be\{\alpha_n p_i, \alpha_n(1 - p_i)\}$ and thus we have that
$E(w_i logw_i) = p_i{\psi_0(\alpha_n p_i + 1) - \psi_0(\alpha_n + 1)}$. Using linearity of expectation
and substituting this expression we obtain the mean. For the variance we follow the properties of the
variance and covariance of sums, which can be expressed from the formulas below:

\begin{align*}
    Var(w_i) = p_i(1 - p_i)/(\alpha_n + 1)
\end{align*}

\begin{align*}
    Cov(w_i, w_j ) = -p_i p_j/(\alpha_n + 1)
\end{align*}

\begin{align*}
Var(w_i,log w_i) = p_i(\alpha_n p_i + 1)\times(\alpha_n + 1)\{\psi_1(\alpha_n p_i + 2) - \psi_1(\alpha_n + 2) + \\
[\psi_0(\alpha_n p_i +2) - \psi_0(\alpha_n + 2)]^2\} - p^2_i \{\psi_0(\alpha_n pi + 1) - \psi_0(\alpha_n + 1)\}^2
\end{align*}

\begin{align*}
Cov(w_i log w_i, w_i) =
p_i(\alpha_n p_i + 1)/(\alpha_n + 1)\{\psi_0(\alpha_n p_i + 2) - \\
\psi_0(\alpha_n + 2)\} - p^2_i\{\psi_0(\alpha_n p_i + 1) - \psi_0(\alpha_n + 1)\}
\end{align*}

\begin{align*}
Cov(w_i log w_i, w_j ) = p_i p_j\{-\psi_0(\alpha_n p_i + 1)/(\alpha_n + 1)+ \\
\psi_0(\alpha_n+1)-\alpha_n \psi_0(\alpha_n + 2)/(\alpha_n + 1)\}
\end{align*}

\begin{align*}
Cov(w_i log w+i, w_j log w_j ) = \alpha_n p_i p_j/(\alpha_n + 1)[\{psi_0(\alpha_n p_i + 1) -\\
\psi_0(\alpha_n + 2)\}\{\psi_0(\alpha_n p_j + 1) -
\psi(\alpha_n + 2)\} - \psi_1(\alpha_n + 2)] -\\
p_i p_j\{\psi(\alpha_n p_i + 1) - \psi_0(\alpha_n + 1)\}\{\psi_0(\alpha_n p_j + 1) - \psi_0(\alpha_n + 1)\}
\end{align*}

So, we get the second part of the result.\\

{\bf Theorem 1 Proof:}\\

For the first case, we have two points $K$ and $K^{*}$ inside the ball.

\begin{align}
\{K_{1,1}+\dots +K_{n,n}=\alpha,K_{i,j}\geq 0,i,j=0,\dots ,n\}\\
\{K^{*}_{1,1}+\dots +K^{*}_{n,n}=\beta,K^{*}_{i,j}\geq 0,i,j=0,\dots ,n\}
\end{align}

Let $\alpha$ and $\beta$ be rational numbers. We have $(1-\rho)^l \times (1+\rho)^m=\frac{\alpha}{b}=\frac{\hat{\alpha}}{\hat{\beta}}=\frac{\hat{\alpha}\times 10^p}{\hat{\beta} \times 10^p}$, with $\hat{\alpha}$ and $\hat{\beta}$ integers. In case $\alpha$ or $\beta$ are irrational numbers then $\frac{\hat{\alpha}}{\hat{\beta}}$ is a very close rational number close to $\frac{\alpha}{\beta}$, so $\frac{\hat{\alpha}}{\hat{\beta}} \simeq \frac{\alpha}{\beta}$, so $\mid \frac{\hat{\alpha}}{\hat{\beta}}- \frac{\alpha}{\beta}\mid\leq \epsilon$ with $\epsilon \rightarrow 0$. Without loss of generality, assume $\hat{\beta}>\hat{\alpha}$, $(1-\rho)^l=\hat{\alpha} \times 10^p$ and $(1+\rho)^m=\hat{\beta} \times 10^p$. We have: $l=\frac{\hat{\alpha} \times 10^p}{ln(1-\rho)}$ and $m=\frac{\hat{\beta} \times 10^p}{ln(1+\rho)}$. With $p \rightarrow \infty$ large we can find integers $l$ and $m$ large enough which satisfy the above equations. So now we have multiplying (14) with $(1-\rho)^l \times (1+\rho)^m=\frac{\alpha}{b}=\frac{\hat{\alpha}}{\hat{\beta}}$:

\begin{align}
\{K_{1,1}+\dots +K_{n,n}=\hat{\alpha},K_{i,j}\geq 0,i,j=0,\dots ,n\}\\
\{K^{*}_{1,1}+\dots +K^{*}_{n,n}=\hat{\alpha},K^{*}_{i,j}\geq 0,i,j=0,\dots ,n\}
\end{align}

We use the first move and we draw $n^2$ independent random samples $y^{*}_{1,1},\dots,y^{*}_{n,n}$ from Gamma distributions each with density:
\begin{align}
\Gamma(\alpha_{i,j}, 1) =\frac{y_{i,j}^{*(\alpha_{i,j}-1)} e^{-y^{*}_{i,j}}}{\Gamma (\alpha_{i,j})}
\end{align}
where, $\alpha_{i,j}$ which denotes the counts in each cell of $K_{i,j}$, and then set
\begin{align}
K_{i,j} = \frac{y^{*}_{i,j}}{\sum_{i,j=1}^{n} y^{*}_{i,j}}\\
K^{*}_{i,j} = \frac{y^{*}_{i,j}}{\sum_{i,j=1}^{n} y^{*}_{i,j}}
\end{align}
If $y^{*}_{i,j}$ are independent $\mathrm{Gamma}(\alpha_{i,j},1)$, for $i,j=1,\dots,n$ then:
\begin{equation}
(K_{1,1},\dots,K_{n,n}) =(K^{*}_{1,1},\dots,K^{*}_{n,n}) =\left
(\frac{y_{1,1}}{\sum_{i,j=1}^{n} y_{i,j}}, \dots, \frac{y_{n,n}}{\sum_{i,j=1}^{n} y_{i,j}} \right) \sim \mathrm{Dirichlet}(\alpha_{1,1},\dots,\alpha_{n,n})
\end{equation}

For the second case, because of the nature of generalized Bayesian bootstrap, any point inside the ball that represent an exchangeable model, with probabilities $w_1,\dots, w_{n^2}$ with $w_i \leq 1$ it is possible to be reached through simulated annealing. Specifically:

\begin{align}
\{w_{1}+\dots +w_{n^2}=1,w_{i}\geq 0,i=0,\dots ,n^2\}
\end{align}

With Bayesian generalized bootstrap we draw $n^2$ independent random samples $y^{*}_{1,1},\dots,y^{*}_{n,n}$ from Gamma distributions each with density:
\begin{align}
\Gamma(\beta_{i,j}, 1) =\frac{y_{i,j}^{*(\beta_{i,j}-1)} e^{-y^{*}_{i,j}}}{\Gamma (\beta_{i,j})}
\end{align}
where, $\beta_{i,j}$ which denotes the counts in each cell of $w_{i,j}$, and then set
\begin{align}
\{w_{1}+\dots +w_{n^2}=1,w_{i}\geq 0,i=0,\dots ,n^2\}
\end{align}
If $y^{*}_{i,j}$ are independent $\mathrm{Gamma}(\alpha_{i,j},1)$, for $i,j=1,\dots,n$ then:
\begin{equation}
(w_1,\dots,w_{n^2}) = =\left
(\frac{y_{1,1}}{\sum_{i,j=1}^{n} y_{i,j}}, \dots, \frac{y_{n,n}}{\sum_{i,j=1}^{n} y_{i,j}} \right) \sim \mathrm{Dirichlet}(\beta_{1,1},\dots,\beta_{n,n})
\end{equation}

\printbibliography

\end{document}